\documentstyle[aps,prb,epsfig]{revtex}

\begin{document}
\twocolumn[\hsize\textwidth\columnwidth\hsize\csname@twocolumnfalse\endcsname
\draft
\title{Josephson Plasma Excitation and Vortex Oscillation Mode
in Josephson Vortes State}
\author{I. Kakeya, T. Wada, and K. Kadowaki}
\address{Institute of Materials Science, University of Tsukuba, Tsukuba, Ibaraki 305-8573 Japan}
\author{M. Machida}
\address{Japan Atomic Energy Research Institute, 2-2-54 Nakameguro, Meguro-ku Tokyo 153-0061, Japan}
\date{\today}
\maketitle

\begin{abstract}
The Josephson plasma resonance has been investigated in 
Bi$_2$Sr$_2$CaCu$_2$O$_{8+\delta}$ single crystals in
parallel magnetic fields to the $ab$-plane.
We found two resonance modes; 
one appears at higher frequency in high fields 
above the plasma frequency $\omega_p$ at zero field and absolute zero, 
and the other lies below $\omega_p$ and is observed only in magnetic fields
without considerable field dependence.
Two resonance lines were also found in numerical simulations in 
a single junction model with randomness of the critical current.
The higher frequency mode is attributed to the Josephson plasma mode
modified by the periodic structure of Josephson vortices,
while the lower frequency mode is interpreted as oscillations of 
Josephson vortices.
\end{abstract}


\vskip.2pc]
\narrowtext

\section{Introduction}
When the external magnetic field is applied parallel to the $ab$-plane of 
high-$T_c$ superconductors such as Bi$_2$Sr$_2$CaCu$_2$O$_{8+\delta}$ (BSCCO), 
Josephson vortices penetrate into block layers.
Since the phase difference between two adjacent CuO$_2$ layers which confine a 
Josephson vortex is $\pi$,
the Josephson vortex is a good example of non-linear dynamics 
in superconductors.
Dynamical properties of Josephson vortices are strongly affected by 
the Josephson plasma 
because the Josephson vortex and the Josephson plasma are solutions of the non-linear and linear part of the sine-Gordon equation~\cite{Cle90}.
Furthermore, a strong emission of electromagnetic wave is expected when the velocity of the JV lattice matches the Josephson plasma 
wave velocity~\cite{Mac00}.
Therefore, the Josephson plasma resonance in the presence of the Josephson vortices is very powerful tool to reveal dynamical nature of the Josephson vortices.

The Josephson plasma resonance (JPR) is known to manifest
the existence of weakly coupled Josephson effects between CuO$_2$ layers
sandwitched by the non-superconducting or insulating block layers.
In particular, in Bi$_2$Sr$_2$CaCu$_2$O$_{8+\delta}$ (BSCCO),
which is a typical high-$T_c$ superconductor and possesses an intrinsic
high anisotropy due to extremely weak Josephson coupling,
the plasma frequency $\omega_p \equiv c/\sqrt{\epsilon}\lambda_c$ 
can be as low as 100 GHz, 
where $\epsilon$ and $\lambda_c$ are the dielectric constant 
and the $c$-axis penetration depth.
Absence of the Landau damping of the quasiparticles in addition
allows us to observe extremely sharp resonance at microwave frequencies
using conventional high sensitive microwave techniques~\cite{Mat95}.
The plasma frequency
in a magnetic field parallel to the $c$-axis
can be expressed as a function of field and temperature:
$\omega_p^2(H,T) = \omega_p^2 \langle \cos \varphi_{l,l+1} (H,T) \rangle$,
where $\varphi_{l,l+1}$ is the gauge invariant phase difference 
between adjacent $l$ and $l+1$-th layers and 
$\langle \cdots \rangle$ denotes spatial and thermal averages\cite{Bul95}.
This means that $\omega_p(H,T)$ can directly sense 
the interlayer phase coherence in a dynamical manner,
which is severely influenced by the thermal fluctuation of vortices.
Thus, $\omega_p$ in general should decrease with increasing field
at a given temperature.
Especially in the vortex liquid state,
$\omega_p$ is expected to follow $H^{-1/2}$ 
as reported previously~\cite{Kos96,Mat97,Shi99a}.

In contrast to this,
the Josephson plasma in fields parallel to the $ab$ plane is 
much more complicated 
because of the non-linear behavior of the Josephson vortex as mentioned above.
In the case of single junctions, the excitation spectrum in the presence of Josephson vortices has been studied by Fetter and Stephen~\cite{Fet68} for example.
They derived an excitation spectrum with two branches:
one is the gapless {\it vortex} branch attributed to the dc sliding mode of 
JVs,
and the other is the {\it plasma} branch identical 
to the transverse Josephson plasma,
which monotonically increases with increasing field.

In a layered system having a multi-stack of 
the intrinsic Josephson junctions like in BSCCO, 
however, physics becomes richer,
because it allows us to observe the longitudinal plasma mode 
(Nambu-Goldstone mode) that
the coherence length comparable to the interlayer distance, 
and the Josephson plasma can propagate not only as the transverse wave 
but also as the longitudinal wave or their mixtures.
Furthermore, the 2D lattice of JVs may be realized 
and the properties may be strongly affected by the intrinsic pinning.
Although much progress of the theories 
for the dynamical properties of JVs~\cite{Mac00,Kos00a}, 
experimental development has been hindered due partly to 
the experimental difficulties restricted by 
precise magnetic field alignment to the $ab$-plane of the crystal,
which requires its extremely high quality. 
Concerning the Josephson plasma in parallel fields,
systematic experimental study 
in order to understand whole spectrum of 
physical properties of the Josephson plasma 
has not been performed, although some preliminary work of JPR 
in parallel fields to the $ab$-plane were reported by 
Matsuda {\it et al.}\cite{Mat97a}.

In this paper,
we present our results of JPR measurements in parallel fields to the $ab$-plane 
as functions of microwave frequency, field intensity, and temperature
in order to reveal the entire nature of the Josephson plasma in 
the presence of JVs.
As a result, we obtained two resonance branches,
which are separated by a finite gap
in good accordance with the recent theoretical approaches.
One observed at higher frequencies is attributed to the Josephson plasma 
coupled with the JV lattice, 
while the other lying at lower frequencies 
is interpreted due to the vortex oscillation mode hardened by residual pinning.

\section{Experiments}
We used an under-doped BSCCO single crystal grown 
by the modified traveling solvent floating zone method. 
The under-doped crystal was 
obtained after annealing at 550 $^{\circ}$C for 12 hours in a vacuum.
The superconducting transition temperature $T_c$ of this crystal was 
determined to be 70.2 K by magnetization measurements with a SQUID magnetometer.
JPR measurements have been performed in a microwave frequency range between 
9.8 and 75 GHz with both reflective and transmission type of 
bridge balance circuits 
with rectangular TE$_{102}$ mode cavity resonators.
The sample was placed inside the cavity 
so that the oscillating electric field $E_{\mathrm{rf}}$ is exerted 
parallel to the $c$-axis to selectively excite the longitudinal plasma 
mode~\cite{PRB98}.
A frequency-stabilized microwave was generated by the signal swept generator 
(HP83650B) or the Gunn oscillators,  
and the resonance was detected as changes of cavity impedance (Q-factor) 
either by field sweep or by temperature sweep. 
The magnetic field was applied by a split-pair of superconducting magnet, 
which can generate horizontal magnetic fields up to 8 T 
and the angle between field direction and crystal axis was changed by
rotating the cavity resonator with respect to  the magnet 
by a precision goniometer.
The field direction exactly parallel to the $ab$-plane was determined by 
looking at
the symmetry of the angular dependence of the resonance within 
an accuracy of 0.01 degree.

\section{Results and Discussion}
\begin{figure}
\begin{center}
\includegraphics[width=\linewidth]{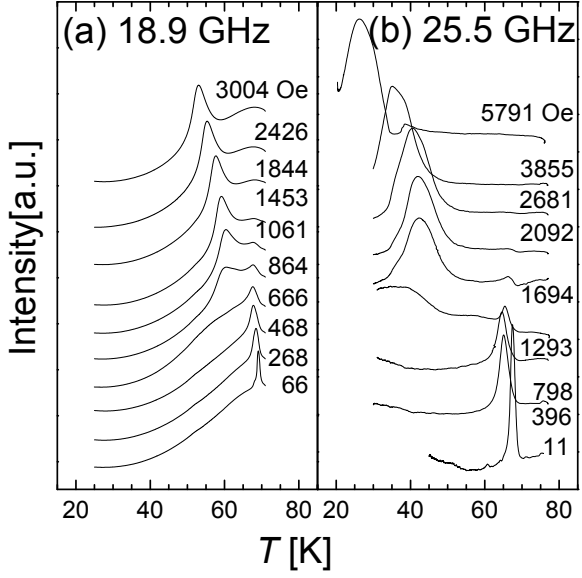}
\end{center}
\caption{
Resonance curves at (a) 18.9 and (b) 25.5 GHz.
Temperature was swept from far above $T_c$ 
to the lowest temperature 
after changing magnetic field, and no hysteresis was found.
}
\label{fig:rawdata1-1}
\end{figure}

Figures \ref{fig:rawdata1-1} (a) and (b) show the resonance curves at 
18.9 and 25.5 GHz, respectively, obtained by sweeping temperature.
Clear two resonance lines without hysteresis at higher and lower temperatures,
and a temperature gap in between the two resonances 
are found at both frequencies.
As the field is increased, 
the resonance at higher temperatures once shifts to lower temperatures,
then turns to shift to higher temperatures above 1 kOe.
It is finally smeared out above 1.5 kOe,
while the resonance at lower temperatures appears at around 1 kOe from
lower temperature side and shifts to lower temperatures with little change 
its line-shape.
Hereafter, we call these resonances at higher and lower temperatures 
HTB (higher temperature branch) and LTB (lower temperature branch), 
respectively.
The fact that two lines are observed and 
the resonance temperature of HTB increases as the field is increased
is peculiar in comparison with the case in fields parallel to the $c$ axis.
It is noted that the LTB resonance obtained by {\it field} sweep was not 
considered here,
because it showed large hysteresis and moreover the line-width was too wide
due to the pinning effect.

\begin{figure}
\begin{center}
\includegraphics[width=\linewidth]{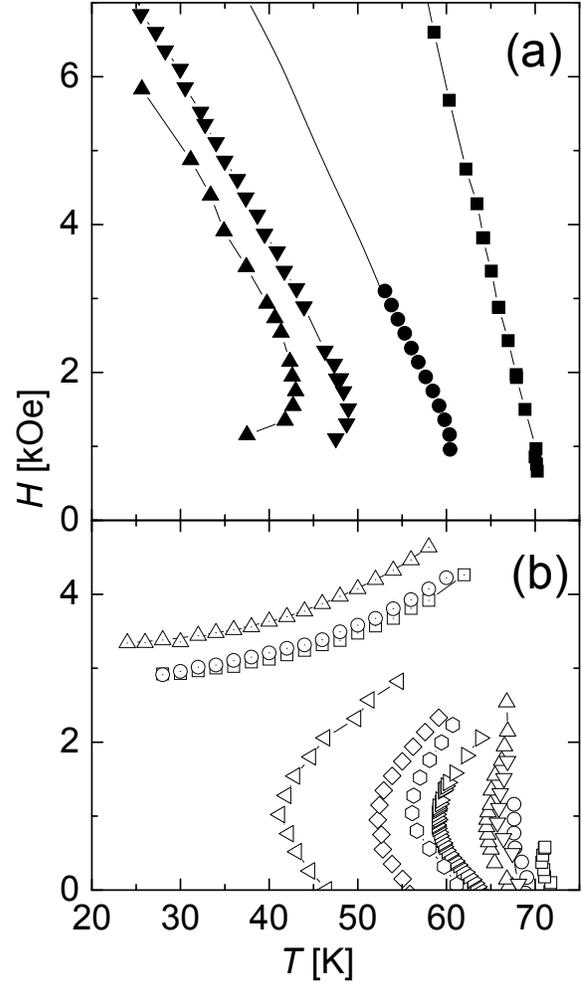}
\end{center}
\caption{
Field--temperature diagrams of the resonance modes for various frequencies.
Solid symbols in (a) indicate LTB obtained by sweeping temperature
at 9.8, 18.8, 22.3, and 25.5 GHz 
and open symbols in (b) indicate HTB 
at 9.8, 18.8, 22.3, 25.5, 34.5, 39.5, 44.2, and 52.3 GHz from right to left.
Open-dotted symbols indicate resonance peaks obtained by sweeping 
field above $\omega_p$ at 61.7, 65.9, and 74.3 GHz from bottom to top.
}
\label{fig:plot1-2} 
\end{figure}

As the frequency is increased, these two resonance lines shift 
to lower temperatures.
Figure \ref{fig:plot1-2} represents the summary of resonance peaks as functions
of field and temperature for various frequencies.
Both HTB and LTB shift to lower temperatures 
and the gap grows rapidly as the frequency increases. 
LTB goes beyond our experimental temperature range above 30 GHz
as shown in the upper panel of Fig. \ref{fig:rawdata1-1},
whereas HTB is obtained at all frequencies.
At a frequency below 60 GHz, 
HTB is observed at zero field at a finite temperature $T_0$, 
where the microwave frequency corresponds to 
the temperature dependent zero-field plasma frequency $\omega_p(T)$.
Extrapolating $\omega_p(T)$ to $T=0$ K, 
the inherent plasma frequency $\omega_p$ can be estimated to be 56.8 GHz.
The turning of HTB is more significant at higher frequencies,
the resonance temperature above 3 kOe exceeds $T_0$ above 30 GHz,
suggesting that the plasma frequency above 3 kOe may exceed $\omega_p$,
whereas LTB just shifts to lower temperatures with increasing frequency.
The resonance peak above $\omega_p$ is also plotted 
in the lower panel of Fig. \ref{fig:plot1-2}.
Similarly to HTB, the resonance line above $\omega_p$ shows no hysteresis in 
field sweep experiments even at low temperatures and peaks by sweeping field 
and temperature correspond well.
Furthermore, the resonance shifts to higher temperatures 
as the field is increased, and is smeared out at higher fields.
Thus, it is natural that this resonance is attributed to HTB observed below $\omega_p$.

\begin{figure}
\begin{center}
\includegraphics[width=\linewidth]{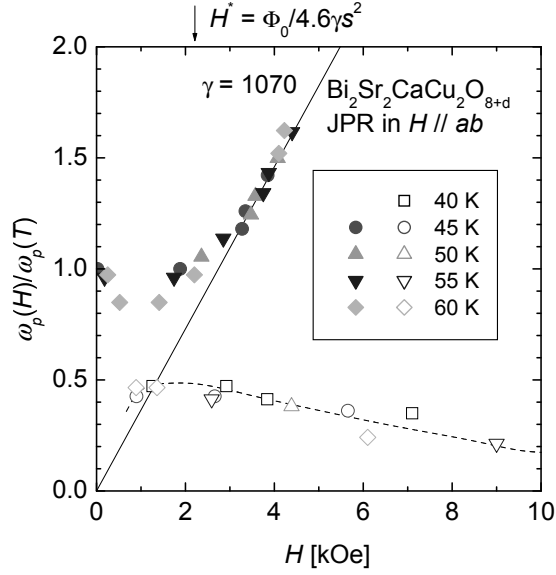}
\end{center}
\caption{
The frequency-field diagram for HTB and LTB.
The plasma frequency $\omega_p(H)$ is normalized 
by the zero-field plasma frequency $\omega_p(T)$.
Solid and open symbols denotes HTB and LTB, respectively.
A solid line is given by Eq. (\ref{ParallelPlasma}) with $\gamma=1070$ 
and a broken curve is a guide for eyes.
}
\label{fig:plot4}
\end{figure}

In order to estimate in-plane field contribution to the Josephson plasma, 
the plasma frequency as a function of magnetic field is 
extracted from Fig. \ref{fig:plot1-2}.
In Fig. \ref{fig:plot4}
the plasma frequencies $\omega_p(H)$ divided by the zero-field 
plasma frequencies $\omega_p(T)$ for various temperatures are shown.
It is clearly seen that two resonance modes 
well-separated by a frequency gap exist: 
the higher one extracted from HTB shows upturn at about 1 kOe, 
then monotonically rises to higher frequencies even above 
$\omega_p$, while the other extracted from LTB shows a gradual decrease with 
increasing field.
It is also found that both are well scaled with $\omega_p(T)$ except for the initial decrease of HTB below 1 kOe.
Since LTB can be observed only in a finite field, 
the existence of JVs would be indispensable for LTB.
At low fields below 0.5 kOe, 
the extrapolation of LTB tends to the zero frequency with decreasing field, 
suggesting that LTB may be a gapless mode at the zero-field limit.
This gives a straightforward interpretation that LTB and HTB 
are the vortex and the plasma branches as suggested 
by Fetter and Stephen~\cite{Fet68}. 

Very recently, 
Koshelev and Machida have developed formulations of 
layered superconductors in high parallel fields to the $ab$-plane 
when a uniform oscillating electric field was applied over the $ab$-plane 
parallel to the $c$-axis~\cite{Mac01}.
Assuming that all block layers are fully occupied by JVs,
which are forced to make a distorted triangular lattice,
they calculated the microwave dissipation due to the Josephson plasma 
modified by oscillations of the JV lattice. 
In this case, the Josephson plasma wave propagates an oblique direction along 
the primitive reciprocal lattice vector of the JV lattice $\bf{q}$, 
which consists of the $c$-axis component $q_c$ fixed on $\pi/s$ 
due to the intrinsic pinning 
and the $ab$-plane component $q_{ab}$ proportional to magnetic field $H$ as 
$2\pi s H/\Phi_0$.
The plasma frequency is determined by the dispersion relation along $\bf{q}$ 
lying between the longitudinal and the transverse plasma modes. 
(Actually, it lies very close to the longitudinal mode 
because of $q_{ab} \ll q_c$.)
As a result, the peak frequency of the dissipation spectrum 
was obtained to be proportional to
the magnetic field parallel to the layers $H$,
\begin{equation}
\omega_p(H)=\omega_p \frac{\pi H \gamma s^2}{\Phi_0}.
\label{ParallelPlasma}
\end{equation}
Assuming $\gamma=1070$, 
which is reasonable for under-doped BSCCO samples, 
this equation gives an excellent agreement with 
the experimental data above 3 kOe at all temperatures
as shown in Fig. \ref{fig:plot4},
where we used $s=15$ {\rm \AA}.
It denotes that the same equation has derived by Bulaeveskii {\it et al}. in 
Ref. \cite{Bul97}.
According to Ichioka,  
all block layers are occupied by JVs above 
$H^*=\Phi_0/4.6\gamma s^2$~\cite{Ich95},
which is obtained to be 2.1 kOe for the above-mentioned $\gamma$.
This is consistent with the fact that the linear field dependence of HTB 
is violated below 3 kOe as shown in Fig. \ref{fig:plot4}.
We therefore conclude that
the linear increase of $\omega_p(H)$ of HTB is attributed to the increase of 
$q_{ab}$ of the JV lattice, 
while $q_c$ is fixed on $\pi/s$ because of the intrinsic pinning.

\begin{figure}[h]
\begin{center}
\includegraphics[width=0.6\linewidth,angle=-90]{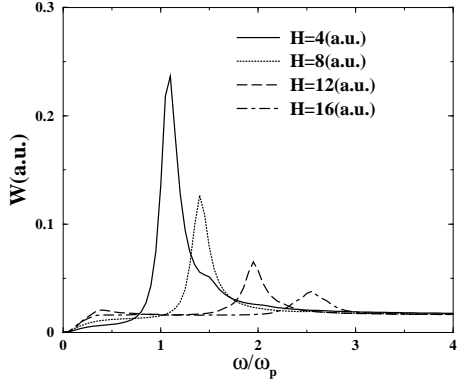}
\end{center}
\caption{
Results of numerical simulations of the microwave absorptions 
in the single junction model with randomness of the critical current 
in four different field intensities.
The amplitude of the randomness is constant and 
the upper-right legend denotes field intensities.
}
\label{sim}
\end{figure}

On the other hand, LTB is considered to originate from the oscillations of 
Josephson vortices, 
which would be hardened by residual pinning.
Figure \ref{sim} represents results of numerical simulations on 
the microwave absorption of the single junction model in the presence of 
randomness of the $c$-axis critical current density.
Introducing the randomness, which represents pinning center of vortices,
a faint peak appears below $\omega_p$.
With increasing field, the faint peak does not change 
while the higher frequency peak
shifts to higher frequencies and its intensity becomes weaker.
This feature is consistent with the experimental results 
in the following points:
the peak of the LTB resonance does not shift to higher frequency and 
the HTB resonance is smeared as field is increased.
However, the intensity of the low-lying resonance is much fainter than
that we observed, 
the intensity of LTB is rather higher than the one of HTB.

Finally, we comment on the initial decrease of HTB resonance frequency
in low fields in connection with the Josephson vortex state.
In the low field region, the effect of the JV lattice is not significant yet,
so that the thermal fluctuation of vortices may play an important contribution 
to suppress the Josephson coupling likewise the case of JPR 
in perpendicular fields.
More detailed analysis of this feature is necessary.

\section{Conclusion}
We have investigated the Josephson plasma mode in 
parallel fields to the $ab$-plane of BSCCO.
We obtained two resonance modes with a temperature dependent gap
in-between.
Strong coupling between Josephson plasma and JV lattice are 
responsible for them.
In high fields where all block layers are occupied with JVs,
one lying higher frequency comes from the Josephson plasma oscillation 
modified by the JV lattice, 
while the other is attributed to the vortex oscillation 
hardened by pinning of Josephson vortex lattice.

\section*{Acknowledgements}
We would like to thank A.E. Koshelev, T. Koyama, S. Takahashi, and M. Tachiki 
for valuable discussions.
This work has been supported by CREST, Japan Science and Technology Cooperation
and is partly supported by the Research Project of The University of Tsukuba.

\bibliographystyle{prsty}
\bibliography{JPJV_ISS}

\end{document}